\documentclass[runningheads]{llncs}

\usepackage[T1]{fontenc}
\usepackage{graphicx}
\usepackage{subcaption}
\usepackage{amsmath}
\usepackage{amssymb}
\usepackage{hyperref}
\usepackage{color}

\usepackage{multirow}
\usepackage{booktabs}
\usepackage{float}
\usepackage[normalem]{ulem}
\usepackage[misc]{ifsym}
\useunder{\uline}{\ul}{}
\hypersetup{
    colorlinks=true,
    linkcolor=blue,
    citecolor=blue,
    urlcolor=blue
}

\begin{document}

%
%\title{Contribution Title\thanks{Supported by organization x.}}
\title{A Unified Framework for Adaptive Representation Enhancement and Inversed 
Learning in Cross-Domain Recommendation}
\titlerunning{AREIL}
\author{Luankang Zhang\inst{1} \and
Hao Wang\inst{1}\textsuperscript{(\Letter)} \and
Suojuan Zhang\inst{2} \and
Mingjia Yin\inst{1} \and
Yongqiang Han\inst{1} \and
Jiaqing Zhang\inst{1} \and
Defu Lian\inst{1} \and
Enhong Chen\inst{1}}
\authorrunning{L. Zhang et al.}
\institute{State Key Laboratory of Cognitive Intelligence, University of Science and Technology of China, Hefei, China\\
\email{\{zhanglk5,mingjia-yin,harley,jiaqing.zhang\}@mail.ustc.edu.cn}\\
\email{\{wanghao3,liandefu,cheneh\}@ustc.edu.cn}
\and
Army Engineering University of PLA, Nanjing, China\\
\email{suojuanzhang@aeu.edu.cn}}
%

%\author{Anonymous Author(s)}
%\institute{}

%\authorrunning{Anon.}
\maketitle              % typeset the header of the contribution
\begin{abstract}

Cross-domain recommendation (CDR), aiming to extract and transfer knowledge across domains, has attracted wide attention for its efficacy in addressing data sparsity and cold-start problems. Despite significant advances in representation disentanglement to capture diverse user preferences, existing methods usually neglect representation enhancement and lack rigorous decoupling constraints, thereby limiting the transfer of relevant information. To this end, we propose a Unified Framework for \textbf{A}daptive \textbf{R}epresentation \textbf{E}nhancement and \textbf{I}nversed \textbf{L}earning in Cross-Domain Recommendation (\textbf{AREIL}). Specifically, we first divide user embeddings into domain-shared and domain-specific components to disentangle mixed user preferences. Then, we incorporate intra-domain and inter-domain information to adaptively enhance the ability of user representations. In particular, we propose a graph convolution module to capture high-order information, and a self-attention module to reveal inter-domain correlations and accomplish adaptive fusion. Next, we adopt domain classifiers and gradient reversal layers to achieve inversed representation learning in a unified framework. Finally, we employ a cross-entropy loss for measuring recommendation performance and jointly optimize the entire framework via multi-task learning. Extensive experiments on multiple datasets validate the substantial improvement in the recommendation performance of AREIL. Moreover, ablation studies and representation visualizations further illustrate the effectiveness of adaptive enhancement and inversed learning in CDR.

%Extensive experiments are conducted to validate the effectiveness of DAFE-CDR, the necessary to perform adaptive enhancement, and the ability to achieve comprehensive disentanglement. 

\end{abstract}

\section{Introduction}

%\begin{figure}
%\centering
%\includegraphics[width=0.9\textwidth,height=0.15\textwidth]{Sections/Figure/Intra.pdf}
%\caption{A toy disentanglement-based CDR scenario. Various shades of the same color indicate different degrees of importance and generality of the features.} \label{intra}
%\end{figure}

%Recently, recommendation systems (RS) have been proposed to capture user interests from historical interactions, with the aim of recommending items to individual users. However, the performance of current recommendation systems is significantly hampered in real-world scenarios due to data sparsity and cold start issues. To this end, cross-domain recommendation (CDR) has been proposed, which seeks to leverage knowledge from source domains to enhance the recommendation performance in target domains~\cite{survey1}.

The recommendation system (RS) has been developed to personalize recommendations by modeling users' preferences based on historical interactions. While attracting wide attention in both academia and industry, RS usually faces data sparsity and cold start issues in real-world applications. To address these challenges, cross-domain recommendation (CDR) has emerged, which improves recommendation performance by transferring knowledge across domains~\cite{survey1}.

CDR aims to model and transfer knowledge across domains, where it is necessary to capture domain-invariant user preferences. Early studies assume that users share consistent preferences across domains, thus directly adapting conventional recommendation methods (matrix factorization~\cite{CMF,CTR}, clustering~\cite{han2023guesr,CBT,Talmud,zhang2022clustering}, and GNN~\cite{PPGN}) to the CDR scenario. Considering diverse user preferences across domains, some studies adopt a two-step paradigm. They first model single-domain preferences using dedicated encoders and then enable cross-domain knowledge transfer by introducing transfer layers. For instance, CoNet~\cite{CoNet} employs Multi-Layer Perceptron (MLP) to encode user preferences and facilitates domain-to-domain information transfer through a dual connection module. BiTGCF~\cite{BiTGCF} adopts a graph-driven framework to model and transfer high-order collaborative information across domains. These approaches ignore the differences in user preferences across domains, potentially leading to the negative transfer problem~\cite{survey2}. Recognizing this, recent research focuses on explicitly separating user preferences into domain-shared and domain-specific components, with the former being transferred across domains. For example, DisenCDR~\cite{DisenCDR} utilizes Variational Autoencoders (VAE) to create embeddings shared across domains and domain-specific embeddings. Additionally, it incorporates two regularizers based on mutual information to supervise the disentanglement process. Similarly, DR-MTCDR~\cite{DRMTCDR} employs a graph neural network to enhance information and disentangle representations with self-supervised learning.

Despite advanced performance achieved by disentanglement-based CDR methods, they exhibit limitations in two crucial aspects:
(1) \textbf{Adaptive Representation Enhancement.} Existing works~\cite{DRMTCDR,CAT-ART,IRCDR,DCCDR} generally assume that domain-shared representations of the same user are completely consistent across domains, leading to an inability to capture diverse user interests. Consequently, they fail to explicitly distinguish the substantial variation in embedding quality caused by imbalanced data distribution and inconsistent domain-shared user preferences across domains. It's a non-trivial challenge to explore the correlations of representations between domains and adaptively transfer important and generalizable knowledge.
(2) \textbf{Inversed Representation Learning.} While several studies employ self-supervised methods like VAE~\cite{DisenCDR} and contrastive learning~\cite{DRMTCDR,GADTCDR} to achieve mutual exclusion of disentangled representations, they cannot guarantee that domain-shared and domain-specific factors are assigned to the corresponding representations respectively. Ideally, these factors should exhibit an inverse relationship, encoding complementary information. Implementing such inversed constraints within a unified framework for learning disentangled user representations remains a critical challenge.

To tackle the aforementioned challenges, in this paper, we propose a Unified Framework for \textbf{A}daptive \textbf{R}epresentation \textbf{E}nhancement and \textbf{I}nversed \textbf{L}earning in Cross-Domain Recommendation, denoted as \textbf{AREIL}.
Specifically, we first initialize item and user representations and then divide user representations into domain-shared and domain-specific components for preference disentanglement.
Second, we enhance user representations through an Adaptive Representation Enhancement Module (AREM). In particular, the intra-domain AREM constructs a user-item interaction graph to capture high-order collaborative information via iterative aggregation. Following this, the inter-domain AREM employs self-attention to evaluate cross-domain relevance and adaptively transfer processed domain-shared embeddings, focusing on important and general components.
Third, we propose the Inversed Representation Learning Module (IRLM) for learning disentangled user preferences in a unified framework, employing domain classifiers and gradient reversal layers (GRL). The GRL reverses the gradient direction to achieve the inversed constraint objective. The domain classifier distinguishes the source of input and generates a comprehensive supervision signal for disentanglement constraints. 
Finally, we leverage a multi-task learning paradigm to optimize the entire framework in an end-to-end manner. In summary, the main contributions of this article can be listed as follows:

\begin{itemize}
    \item[$\bullet$] We study the dual-target cross-domain recommendation problem from a novel perspective, which focuses on adaptive representation enhancement and inversed learning for user preferences disentanglement.
    \item[$\bullet$] To enhance user representations, we propose an adaptive representation enhancement module. This module enables the exploration of inter-domain relevance, allowing for the adaptive transfer of important and general factors that contribute to improved recommendation performance.
    \item[$\bullet$] We leverage domain classifiers and gradient reversal layers within the inversed representation learning module to constrain the disentanglement of user preferences in a unified framework.
    %We design CFDM for comprehensive disentanglement of user interests. This module enables the introduction of high-quality self-supervised signals into the decoupling process.
    \item[$\bullet$] Extensive experiments confirm AREIL consistently outperforms state-of-the-art models. Supplementary ablation studies and representation visualizations further indicate our module can learn more informative and precise disentangled user representations in the cross-domain recommendation.
    
    %Extensive experiments have been conducted to validate the effectiveness of DAFE-CDR, which has consistently outperformed current state-of-the-art models. Additionally, supplementary ablation studies and embedding visualizations further confirm the efficacy of feature enhancement and the quality of disentanglement.
\end{itemize}

\section{Related Work}

\subsection{Cross-Domain Recommendation}

Recommendation systems~\cite{END4Rec,han2023guesr,wang2021hypersorec,wu2023survey,yin2023apgl4sr} typically face challenges associated with data sparsity and cold-start scenarios. Cross-domain recommendation (CDR) has emerged as a promising approach that leverages knowledge from a source domain to improve recommendation performance in a target domain~\cite{survey1}.

Within the field of CDR, the central focus is on modeling domain-invariant user preferences and facilitating information transfer across diverse domains. Early approaches primarily extended single-domain recommendation methods, assuming that users share the same interests across different domains. For instance, techniques such as matrix factorization~\cite{BPR}, graph-based methods~\cite{LightGCN,wang2019mcne,NGCF}, and contrastive learning~\cite{DirectAU,catcl,yi2023contrastive,SimGCL} are applied to individual domains, employing a basic transfer module to integrate information across domains. Although these modules outperform single-domain recommendation methods in performance, they disregard variations in user interests across domains and give rise to the negative transfer problem~\cite{survey2}. Some methods utilize two separate encoders to model user interests and introduce more complex transfer modules. For example, CoNet~\cite{CoNet} introduced dual connections in a Multi-Layer Perceptron (MLP) network to achieve deep bidirectional knowledge migration. DARec~\cite{DARec} utilized autoencoders and adversarial learning for knowledge transfer among shared users. DDTCDR~\cite{DDTCDR} introduced a deep dual transmission network with cross-domain implicit orthogonal mapping to preserve the similarity of user preferences across domains. Furthermore, subsequent research introduced Graph Neural Network (GNN)-based transfer modules, incorporating higher-order information into the transfer process. For example, PPGN~\cite{PPGN} captures the multi-hop propagation of user preferences, explicitly models cross-domain interactions and preserves the structural information. BiTGCF~\cite{BiTGCF} leverages higher-order connectivity in single-domain user-item graphs through a novel feature propagation layer, facilitating bidirectional knowledge exchange between two domains by utilizing overlapped users as intermediaries. Despite advancements in encoder architectures, existing methods fail to adequately capture the diversity of user preferences across domains, leading to suboptimal recommendation performance.
%Different from these previous works, we enhance features adaptively, highlighting important and general information.

% With the emergence and development of deep learning in recent years, many deep learning-based approaches~\cite{CoNet,DARec,DDTCDR,AAM,PPGN,BiTGCF} have been proposed and significantly improved the performance of CDR. CoNet~\cite{CoNet}, proposed by Hu et al., introduced dual connections in a multilayer feedforward network to accomplish deep bidirectional knowledge migration. The DARec model~\cite{DARec} leverages autoencoders and adversarial learning for knowledge transfer among shared users. DDTCDR~\cite{DDTCDR} proposes a deep dual transmission network that employs cross-domain implicit orthogonal mapping to preserve similarity between user preferences across domains and efficiently compute the inverse mapping. Similarly, AAM~\cite{AAM} integrates autoencoders and self-attention mechanisms for cross-domain migration. PPGN~\cite{PPGN} captures multi-hop propagation of user preferences, explicitly modeling cross-domain interactions and preserving structural information. BiTGCF~\cite{BiTGCF} capitalizes on higher-order connectivity in single-domain user-item graphs through a novel feature propagation layer, enabling bi-directional knowledge transfer between two domains by utilizing regular users as bridges. Deep cross-domain recommendation methods exhibit stronger feature mining capabilities than shallow methods, leading to improved cross-domain recommendation performance.

\subsection{Disentanglement Learning in CDR}

To further alleviate the negative transfer problem, disentanglement-based CDR methods~\cite{disentangle4,wang2021decoupled,disentangle3,disentangle1,disentangle2} have gained prominence. These methods disentangle features into domain-specific and domain-shared components and solely migrate domain-shared information, which mitigates potential interference resulting from the migration of domain-specific information. Variational AutoEncoder (VAE)~\cite{beta_vae} is known to be effective in disentangling, motivating works like DisenCDR~\cite{DisenCDR} and MTNet~\cite{MTNet} to disentangle features with different disentanglement objectives. Adversarial learning has also been introduced to CDR. For example, DA-CDR~\cite{DA-CDR} controlled the decoupling process with orthogonality constraints, while DIDA-CDR~\cite{DIDA-CDR} introduced the concept of ``domain-independent information'', challenging the accuracy of encoding domain-shared representations and achieves more precise decoupling. Besides, some works~\cite{DRMTCDR,SCD,DCCDR} proposed to disentangle representations in a Self-Supervised Learning fashion, which provided valuable guidance for disentanglement in the context of data sparsity. However, incomplete disentanglement in these methods, due to either limited decoupling control or insufficient mutual exclusion, results in inadvertent feature transfer across domains and reduced recommendation performance.
%In contrast to existing methods that only partially distinguish user interests, we aim to attain complete feature disentanglement.

%While existing methods partially achieve feature decoupling and mitigate the negative migration problem, ensuring accurate differentiation between highly coupled domain-shared and domain-specific information, as well as guaranteeing the inclusion of sufficient information in the encoded representations, remains challenging. To tackle these issues, our proposed approach introduces adversarial signaling to better distinguish between domain-shared and domain-specific information while controlling their decoupling process. Furthermore, we enhance the decoupling effect by progressively injecting domain-shared information for each node individually. This refinement contributes to improved decoupling outcomes and overall recommendation performance.

\section{Problem Definition}

In this section, we will provide formal definitions to facilitate a more precise explanation of the problem. %Specifically, we will begin by introducing the task formulation for the cross-domain recommendation, which enhances features by transferring information across domains. Furthermore, we will introduce the concept of feature disentanglement, which seeks to model varying user preferences within different domains.
Given two domains, denoted as $X$ and $Y$, CDR leverages information from a relatively richer domain to improve performance in a sparser domain. This is achieved by capturing domain-invariant user preferences and transferring information across domains. In the following, we mainly focus on the
dual-target CDR. The formalization of the problem is as follows:

\textbf{Definition (Dual-target Cross-Domain Recommendation)}. \textit{We consider a general scenario wherein users are fully overlapped while items remain non-overlapped. Let $\mathcal{D}^X=(\mathcal{U}, \mathcal{V}^X, \mathcal{E}^X)$ and $\mathcal{D}^Y=(\mathcal{U}, \mathcal{V}^Y, \mathcal{E}^Y)$ denote the interaction data of domain $X$ and $Y$ respectively. Specifically, $\mathcal{U}$ represents the shared user set between the two domains, with a size ${|\mathcal{U}|}$. Furthermore, $\mathcal{V}^X$ and $\mathcal{V}^Y$ indicate the non-overlapping item sets specific to domain $X$ and domain $Y$, respectively. The interactions between users and items are captured using the edge sets $\mathcal{E}^X$ and $\mathcal{E}^Y$. Moreover, binary interaction matrices $A^X \in\{0,1\}^{|\mathcal{U}| \times|\mathcal{V}^X|}$ and $A^Y \in\{0,1\}^{|\mathcal{U}| \times|\mathcal{V}^Y|}$ are employed to encode the user-item interactions in domains $X$ and $Y$, where $A_{i j}=1$ indicates an interaction between user $u_i \in \mathcal{U}$ and item $v_j \in \mathcal{V}$, while $A_{i j}=0$ indicates the absence of such interaction. Then, given $(\mathcal{D}^X, \mathcal{D}^Y, A^X, A^Y)$, The dual-target CDR aims to improve the performance of both domains $X$ and $Y$ simultaneously.}

Traditional dual-target CDR methods neglect the variation in user preferences across domains, thereby limiting the efficacy of knowledge transfer. To capture diverse user preferences, there arises a necessity to disentangle user interests, a process that can be precisely formalized as follows:

\textbf{Definition (User Preferences Disentanglement)}. \textit{Given the user representation $\mathbf{Z}_u^X$ in domain $X$, we disentangle it into domain-shared $\mathbf{Z}_{\text{u,sha}}^X$ and domain-specific components $\mathbf{Z}_{\text{\textit{u,spe}}}^X$, respectively. Similarly, we disentangle the user representation $\mathbf{Z}_u^Y$ in domain $Y$ into $\mathbf{Z}_{\text{\textit{u,sha}}}^Y$ and $\mathbf{Z}_{\text{\textit{u,spe}}}^Y$. The disentangled representation is expected to encode corresponding information, wherein domain-shared encoding denotes domain-invariant information, and domain-specific encoding conveys domain-related information.}

Current disentanglement methods overlook representation enhancement and lack sufficient decoupling constraints. We present an end-to-end framework for addressing these issues. The methodological details are fully described in Sec.~\ref{sec:Methodology}.

\section{Methodology}\label{sec:Methodology}
\begin{figure*}[t]
\centering
\includegraphics[width=\textwidth, height=0.44\textwidth]{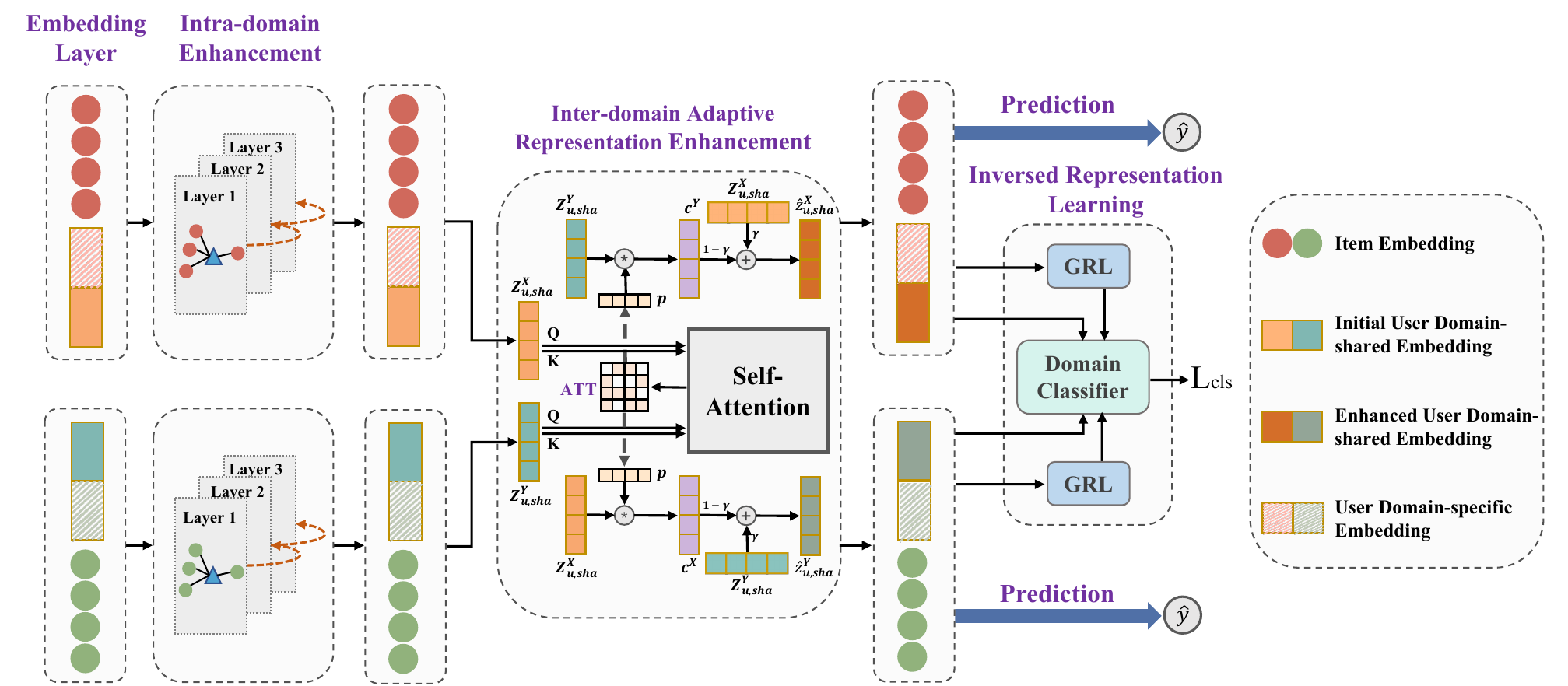}

\caption{The AREIL framework for adaptive representation enhancement and inversed representation learning.
%The framework of AREIL. We achieve adaptive feature enhancement and comprehensive disentanglement.
%Our approach comprises the subsequent components: (1) Disentanglement-based Embedding Layer, initializing representations for users and items in both domains; (2) Feature Propagation Layer, utilizing LightGCN on the user-item interaction graph to extract higher-order collaborative information; (3) Domain Adaptive Fusion Module, facilitating explicit bidirectional knowledge transfer between domains adaptively at the node level; (4) Feature Disentanglement Module, incorporating self-supervised signals to supervise the decoupling process; and (5) Prediction Layer.
} \label{overall}
\end{figure*}

%\subsection{Overall Framework}
%Despite notable advancements in current Cross-Domain Recommendation Systems (CDRs), limited attention has been given to the disentanglement of user interests. To this end,

In this section, we introduce a unified framework for \textbf{A}daptive \textbf{R}epresentation \textbf{E}nhancement and \textbf{I}nversed \textbf{L}earning in Cross-Domain Recommendation (AREIL) to achieve adaptive representation enhancement and inversed representation learning. The framework is illustrated in Fig.~\ref{overall}. In Sec.~\ref{sec:Embedding}, AREIL initially disentangles mixed user preferences into domain-shared and domain-specific components by splitting user representations equally. Moving forward to Sec.~\ref{sec:Enhancement}, the Adaptive Representation Enhancement Module (AREM) is designed to enhance the modeling of user preferences. In particular, intra-domain high-order information is incorporated by utilizing LightGCN to aggregate neighbor node representations, and the inter-domain AREM adaptively facilitates the transfer of important and general factors across domains by self-attention. Additionally, Sec.~\ref{sec:FDM} introduces domain classifiers and gradient reversal layers to learn disentangled user representations in a unified framework by generating effective self-supervised signals. Finally, we make predictions and optimize the entire framework using a multi-task learning paradigm. %We will elaborate on the technical details of each module.

%In Sec.~\ref{sec:Embedding}, we investigate the varying influence and quality of domain-shared features across different domains. Inspired by this investigation, we introduce a novel concept of disentangled user embeddings, capable of encoding varying degrees of influence and trustworthiness. This representation paradigm serves as the foundational basis for our model. Moving on to Sec.~\ref{sec:Propagation}, we apply LightGCN to the user-item interaction graph to extract higher-order collaborative information, thereby enhancing representation capabilities. Following this, in Sec.~\ref{sec:DAF}, we explore adaptive feature fusion, highlighting influential and trustworthy information. We introduce the Domain Adaptive Fusion module (DAF) to selectively incorporate features and explicitly integrate information from another domain on the node level, forming the core of the disentanglement process. Additionally, in Sec.~\ref{sec:FDM}, we delve into comprehensive disentanglement. Inspired by adversarial learning, we propose the Feature Disentanglement Module (FDM). FDM employs domain classifiers and gradient inversion layers to effectively generate self-supervised signals, facilitating mutual exclusivity and encoding constraints. Finally, we make predictions and optimize the entire framework using a multi-task learning paradigm. 

\subsection{Disentanglement-based Embedding Layer}\label{sec:Embedding}
% The embedding layer is responsible for encoding users and items into the feature space. 
In traditional cross-domain recommendation methods~\cite{CoNet,DARec}, it is assumed that users maintain consistent interests across different domains, which often results in negative transfer because it fails to effectively capture diverse user preferences. To overcome this limitation, we introduce a Disentanglement-based Embedding Layer that segregates user preferences into two components: domain-shared and domain-specific. Specifically, considering domain $X$, we evenly partition $\mathbf{Z}_u^X \in \mathbb{R}^{|\mathcal{U}| \times d}$ into two components: domain-shared user embeddings $\mathbf{Z}_{\text{\textit{\textit{u,sha}}}}^X \in \mathbb{R}^{|\mathcal{U}| \times d/2}$ and domain-specific user embeddings $\mathbf{Z}_{\text{\textit{\textit{u,spe}}}}^X \in \mathbb{R}^{|\mathcal{U}| \times d/2}$, i.e. $\mathbf{Z}_u^X=\mathbf{Z}_{\text{\textit{\textit{u,sha}}}}^X \| \mathbf{Z}_{\text{\textit{\textit{u,spe}}}}^X$. The even number $d$ denotes the feature dimension. The Disentanglement-based Embedding Layer serves as the foundational basis for our model, defining domain-shared preferences separately for each domain. This design enables the embeddings to express the varying importance and generality of domain-independent features within each domain. The same process is applied to domain $Y$. Moreover, we utilize $\mathbf{Z}_v^X \in \mathbb{R}^{|\mathcal{V}^X| \times d}$ and $\mathbf{Z}_v^Y \in \mathbb{R}^{|\mathcal{V}^Y| \times d}$ to represent item sets.

\subsection{Adaptive Representation Enhancement Module}\label{sec:Enhancement}
%After encoding users and items, we apply feature propagation techniques to aggregate structural information and enhance the representations. Specifically, we leverage user-item interactions to construct a heterogeneous bipartite graph and utilize GCN~\cite{GCN} to extract higher-order collaborative signals from graph structures. Node embeddings are refined through the aggregation of messages from adjacent nodes. This procedure can be summarized as follows:
Given the disentangled embeddings, we employ the Adaptive Representation Enhancement Module (AREM) to enhance their ability by incorporating both intra-domain and inter-domain information.

%\subsubsection{Intra-domain Adaptive Feature Enhancement Module.}

\subsubsection{Intra-domain Enhancement.}
%We employ feature propagation techniques to capture high-order interaction between users and items, enhancing representations through intra-domain information. Specifically, we leverage user-item interactions to construct a heterogeneous bipartite graph and apply GCN~\cite{GCN} to extract higher-order collaborative signals from the graph structures. Node embeddings are refined through the aggregation of messages from adjacent nodes. This procedure can be summarized as follows:

%\begin{equation}
%e_u^{(k+1)}=AGG(e_u^{(k)}, e_i^{(k)}: i \in \mathcal{N}_u),
%\end{equation}
%where $k$ represents the current graph convolutional layer, $e_u^{(k)}$ refers to the target node embedding in the $k$-th layer, $\mathcal{N}_u$ represents the set of neighboring nodes for the target node $u$, $e_i^{(k)}$ denotes the $k$-th layer embedding of the neighboring nodes $i$, and $AGG$ denotes an aggregation function that combines the $k$-th layer embedding of the target node and its neighboring nodes to obtain the target node embedding $e_u^{(k+1)}$ in the ($k$+1)-th layer.

To capture high-order collaborative information, we construct a heterogeneous bipartite graph by leveraging user-item interactions. Subsequently, we apply LightGCN~\cite{LightGCN} to enhance node representations through the aggregation of embeddings from adjacent nodes. In the context of domain $X$, the linear propagation procedure can be summarized as follows:

%In particular, we employ LightGCN~\cite{LightGCN}, a streamlined linear model well-suited for recommendation systems, utilizing neighborhood aggregation for collaborative filtering. It learns user and item embeddings through linear propagation across the user-item interaction graph. For instance, in the context of domain $X$, the user and item embeddings in the ($k$+1)-th layer are expressed as follows:
\begin{equation}
%\begin{gathered}
\mathbf{Z}_u^{X,k+1}=\sum_{i \in \mathcal{N}_u^X} \frac{1}{\sqrt{\left|\mathcal{N}_u^X\right|\left|\mathcal{N}_i^X\right|}} \mathbf{Z}_i^{X,k},\\
%\quad \mathbf{Z}_i^{X,k+1}=\sum_{u \in \mathcal{N}_i^X} \frac{1}{\sqrt{\left|\mathcal{N}_i^X\right|\left|\mathcal{N}_u^X\right|}} \mathbf{Z}_u^{X,k}.
%\end{gathered}
\end{equation}
where $k$ represents the current graph convolutional layer, $\mathcal{N}_u$ represents the set of neighboring nodes in domain $X$ for the target node $u$, and $\mathbf{Z}_i^{X,k}$ denotes the embedding of the node $i$ in $k$-th layer.

To fuse distinct semantic information, we concatenate the representations obtained from different $K$ layers, resulting in enhanced node representation $\mathbf{Z}_u^X$:
\begin{equation}
\mathbf{Z}_u^X=\mathbf{Z}_u^{X,0}\|\ldots\| \mathbf{Z}_u^{X,K}.%, \quad \mathbf{Z}_i^X=\mathbf{Z}_{i}^{X,0}\|\ldots\| \mathbf{Z}_{i}^{X,K}.
\end{equation}
%\subsubsection{Inter-domain Adaptive Feature Enhancement Module.}

\subsubsection{Inter-domain Enhancement.}

%After incorporating intra-domain information into the embeddings, we initiate the process of inter-domain enhancement to integrate information across domains. We account for the unique characteristics inherent in cross-domain recommendation systems compared to more general scenarios. Firstly, CDR faces significant inter-domain disparities, necessitating powerful strategies for cross-domain information sharing. Additionally, CDR often encounters data skewness, where only a few users possess abundant interaction records, leading to substantial variation in representation quality. Furthermore, preferences that exist simultaneously in both domains can exhibit distinct behaviors. To align our method with these unique characteristics, addressing domain disparity and adaptively emphasizing important and generality information, we introduce an Inter-domain Adaptive Feature Enhancement Module (Inter-domain AREM). This module aims to fuse information across domains adaptively, selectively incorporating features and explicitly integrating information from another domain on the user level, forming the foundation of the disentanglement process.

To address significant data sparsity issues in CDR, we further enhance representations by transferring information across domains. Unfortunately, CDR frequently encounters significant variations in representation quality due to data skewness, where a few users contribute to most interaction records. Additionally, preferences existing simultaneously in both domains may exhibit distinct behaviors. These characteristics make it impractical to simply transfer domain-shared representations. To address these issues, we introduce the Inter-domain Adaptive Representation Enhancement Module (Inter-domain AREM). This module employs a self-attention mechanism~\cite{Transformer} to explore correlations between domains and adaptively transfer important and general knowledge at the node level. For simplicity, we will illustrate the procedure using domain $X$, and a similar process can be extended to domain $Y$.

To illustrate the inherent relationship between domain $X$ and domain $Y$, we introduce the attention matrix $ATT^X$. Specifically, given the domain-shared user representations $\mathbf{Z}_{\text{\textit{\textit{u,sha}}}}^X \in \mathbb{R}^{|\mathcal{U}| \times d/2}$ from domain $X$ and $\mathbf{Z}_{\text{\textit{\textit{u,sha}}}}^Y \in \mathbb{R}^{|\mathcal{U}| \times d/2}$ from domain $Y$, we consider them as sequences of length 2. These sequences are combined to form $\mathbf{Z}_{\text{\textit{u}}} \in \mathbb{R}^{|\mathcal{U}| \times d}$, i.e. $\mathbf{Z}_{\text{\textit{u}}}=\mathbf{Z}_{\text{\textit{u,sha}}}^X \| \mathbf{Z}_{\text{\textit{u,sha}}}^Y$. Next, we introduce the learnable query weight matrix $W_Q^X\in \mathbb{R}^{d \times d}$ and key weight matrix $W_K^X\in \mathbb{R}^{d \times d}$. We then obtain query vector $\mathbf{Q}^X$ and key vector $\mathbf{K}^X$:
\begin{equation}
    \mathbf{Q}^X = \mathbf{Z}_{\text{\textit{u}}} W_Q^X, \quad \mathbf{K}^X = \mathbf{Z}_{\text{\textit{u}}} W_K^X.
\end{equation}

Therefore, we get the attention matrix $ATT^X$ by calculating the dot product between the query vector $\mathbf{Q}^X$ and the key vector $\mathbf{K}^X$, i.e. $ATT^X = (\mathbf{Q}^X)^T\mathbf{K}^X$.
%Subsequently, by computing the dot product between the query vector $\mathbf{Q}^X$ and the key vector $\mathbf{K}^X$, we derive the attention matrix $ATT^X = (\mathbf{Q}^X)^T\mathbf{K}^X$. This attention matrix $ATT^X$ showcases the common relationship between domain $X$ and domain $Y$.

%Subsequently, by computing the dot product between the query vector $\mathbf{Q}^X$ and the key vector $\mathbf{K}^X$, we derive the attention matrix $ATT^X = (\mathbf{Q}^X)^T\mathbf{K}^X$. This attention matrix $ATT^X$ showcases the common relationship between domain $X$ and domain $Y$.

Based on the attention matrix $ATT^X$, we introduce a gating mechanism to emphasize the important and general components in $\mathbf{Z}_{\text{\textit{u,sha}}}^Y$. Specifically, we calculate the feature-related distribution $p$ by summing the attention matrix $ATT^X$ along the query dimensions. A higher value of $p$ indicates a stronger correlation between the corresponding elements in $\mathbf{Z}_{\text{\textit{u,sha}}}^X$ and $\mathbf{Z}_{\text{\textit{u,sha}}}^Y$, suggesting greater importance for injection into domain $X$. Next, utilizing $p$, we emphasize important and general components in $\mathbf{Z}_{\text{\textit{u,sha}}}^Y$ and derive the commonality $c^Y$:
\begin{equation}
c^Y=\mathbf{Z}_{\text{\textit{u,sha}}}^Y \odot \operatorname{Norm}(p),
\end{equation}
where $\odot$ signifies element-wise product, and $\operatorname{Norm}(\cdot)$ denotes normalization.

Finally, we derive the enhanced domain-shared representation $\mathbf{\hat{Z}}_{\text{\textit{u,sha}}}^X$ of domain $X$ by integrating commonality in domain $Y$:
\begin{equation}
    \mathbf{\hat{Z}}_{\text{\textit{u,sha}}}^X=\gamma_s \mathbf{Z}_{\text{\textit{u,sha}}}^X+(1-\gamma_s) c^Y,
\end{equation}
where $\gamma_s$ is the weight score that balances the domain-shared user representation $\mathbf{Z}_{\text{\textit{u,sha}}}^X$ in domain $X$ and commonality $c^Y$ in domain $Y$.
%Utilizing both the self-attention and gating mechanisms, we adaptively integrate information across domains, thereby establishing a robust groundwork for subsequent feature disentanglement.

%effectively integrate the commonality information from domain $Y$ into the domain-sharing representation of domain $X$, facilitating more informative knowledge transfer.

\subsection{Inversed Representation Learning Module}\label{sec:FDM}
%After enhancing the embeddings adaptively, we introduce a constraint aimed at achieving comprehensive disentanglement. Previous investigations~\cite{DisenCDR,SEM-MacridVAE} have been deficient in high-quality self-supervised signals, primarily concentrating on achieving mutual exclusivity between domain-shared and domain-specific embeddings. In other words, there has been a neglect in enforcing these embeddings to encode accurate information. Consequently, the issue of constraining domain-shared and domain-specific embeddings to encode domain-invariant and domain-dependent information, respectively, has not been thoroughly explored. Drawing inspiration from adversarial learning, we present the  Inversed Representation Learnin Module (IRLM), which incorporates a domain classifier and a gradient inversion layer to provide supervision during the decoupling process, thereby achieving comprehensive disentanglement.

Building upon adaptively enhanced representations, we establish a unified framework for learning disentangled user preferences. Existing studies~\cite{DisenCDR,SEM-MacridVAE} suffer from limitations in both self-supervised signal quality and fail to enforce domain-specific and domain-shared representations for encoding domain-dependent and domain-invariant information, respectively. To this end, we introduce the Inversed Representation Learning Module (IRLM). This module integrates a domain classifier and a gradient reversal layer to supervise the disentanglement process within a unified framework. 

The domain classifier acts as a crucial tool that takes in user representations and predicts the domains from which they originate. For domain-specific user representations, we utilize them to encode domain-dependent information, allowing the domain classifier to readily discern the origin of inputs. The constraint is accomplished through the minimization of the loss function:
%Specifically, our aim is to employ domain-specific embeddings to encode domain-dependent information, enabling the domain classifier to readily distinguish the origin of inputs by minimizing the loss function:
\begin{equation}
    %\mathcal{L}_{cls_{spe}}=\ell(DC(\mathbf{\hat{Z}}_{\text{\textit{u,spe}}}^X), O_{\text{\textit{spe}}}^X) + \ell(DC(\mathbf{\hat{Z}}_{\text{\textit{u,spe}}}^Y), O_{\text{\textit{spe}}}^Y).
    \mathcal{L}_{cls_{spe}}=\ell(DC(\mathbf{Z}_{\text{\textit{u,spe}}}^X), O) + \ell(DC(\mathbf{Z}_{\text{\textit{u,spe}}}^Y), O),
\end{equation}
where $l$ represents the cross-entropy loss, $O$ denotes the domain label ($O = 0$ for domain $X$ and $O = 1$ otherwise), and $DC(\cdot)$ denotes the output of the domain classifier, which is implemented as a multi-layer perceptron (MLP).

Concerning domain-shared representations, we intend for them to encode domain-independent information and confuse the domain classifier. In other words, the objective of domain-shared user representations is to maximize the domain classification error, directly contrasting the goal of the domain classifier, which is to precisely categorize input representations. To learn these two inversed optimization objectives in a unified framework, we design a gradient reversal layer (GRL). The GRL automatically reverses the gradient direction during backpropagation and remains constant during forward propagation:

%In contrast, we intend for the domain-shared representation to encode domain-independent information and confuse the domain classifier. In other words, during training, we simultaneously pursue two objectives: (1) enabling the classifier to accurately classify input representations and (2) maximizing the error in domain classification by making the shared representations of domains $X$ and $Y$ as indistinguishable as possible. Inspired by Domain Adaptation~\cite{DANN}, we utilize a gradient reversal layer (GRL) to achieve this adversarial training. The GRL automatically reverses the gradient direction during backpropagation, creating a consistent transformation in the forward propagation:
\begin{equation}
    G R L(x)=x, \quad \frac{d G R L}{d x}=-\lambda I,
\end{equation}
where $I$ denotes the unit matrix, and $\lambda$ is a dynamically adjusted parameter that balances the relationship between domain adaptation and classification accuracy.

Through the incorporation of GRL, we compel the domain-shared representations to encode invariant user preferences by minimizing the loss:
%Similarly, we enforce the domain sharing to represent encoded domain invariant user preferences by minimizing the loss function $\mathcal{L}_{cls_{sha}}$, which is analogous to $\mathcal{L}_{cls_{spe}}$ but incorporating an additional GRL transformation:
\begin{equation}
    %\mathcal{L}_{cls_{sha}}=\ell(DC(GRL(\mathbf{\hat{Z}}_{\text{\textit{u,sha}}}^X)), O_{\text{\textit{sha}}}) + \ell(DC(GRL(\mathbf{\hat{Z}}_{\text{\textit{u,sha}}}^Y)), O_{\text{\textit{sha}}}).
    \mathcal{L}_{cls_{sha}}=\ell(DC(GRL(\mathbf{\hat{Z}}_{\text{\textit{u,sha}}}^X)), O) + \ell(DC(GRL(\mathbf{\hat{Z}}_{\text{\textit{u,sha}}}^Y)), O).
\end{equation}

The disentanglement of user representations is rigorously constrained by final classification loss $\mathcal{L}_{cls}$, which arises from the fusion of domain-shared and domain-specific components:
\begin{equation}
    \mathcal{L}_{cls}=\mathcal{L}_{cls_{sha}}+\mathcal{L}_{cls_{spe}}.
\end{equation}

\subsection{Prediction Layer and Multi-task Learning}
%Following the adaptive enhancement of features and the attainment of comprehensive disentanglement, we can acquire user embeddings $\mathbf{Z}_u^X=\mathbf{Z}_{\text{\textit{u,sha}}}^X\| \mathbf{Z}_{\text{\textit{u,spe}}}^X$, $\mathbf{Z}_u^Y=\mathbf{Z}_{\text{\textit{u,sha}}}^Y\| \mathbf{Z}_{\text{\textit{u,spe}}}^Y$, and item representations $\mathbf{Z}_v^X$, $\mathbf{Z}_v^Y$. The prediction scores $\hat{r}_{u i}^{\mathrm{X}}$ and  $\hat{r}_{u i}^{\mathrm{Y}}$ are computed as the dot product between the user and item representations:
Following the adaptive enhancement of representations and the inversed learning of disentangled user preferences, we can acquire the prediction scores $\hat{r}_{u i}^{\mathrm{X}}$ and $\hat{r}_{u i}^{\mathrm{Y}}$:
\begin{equation}
    \hat{r}_{u i}^{\mathrm{X}}={\mathbf{Z}_u^X}^T \mathbf{Z}_i^X, \quad \hat{r}_{u i}^{\mathrm{Y}}={\mathbf{Z}_u^Y}^T \mathbf{Z}_i^Y.
\end{equation}
We then adopt a cross-entropy function to measure the model's performance:
\begin{equation}
    \mathcal{L}_{rec}=-\sum r_{u i} \log \hat{r}_{u i}+(1-r_{u i}) \log (1-\hat{r}_{u i}).
\end{equation}
%To prevent overfitting, we apply $L_2$ regularization to the parameter set $\Theta$, resulting in the regularized loss $\mathcal{L}_{reg}=\|\Theta\|_2^2$. Hence, the overall loss function is:
To prevent overfitting, the regularization loss $\mathcal{L}_{reg}=\|\Theta\|_2^2$ is applied to the parameter set $\Theta$. Consequently, the joint learning objective function of AREIL could be defined as the following multi-task learning framework:
%\begin{equation}
%    \mathcal{L}_{reg}=\|\Theta\|_2^2,
%\end{equation}
%where $\Theta$ represents the parameter set. 
%Consequently, the overall loss function is formulated as 
\begin{equation}
    \mathcal{L}=\mathcal{L}_{rec}+\lambda_1\mathcal{L}_{cls}+\lambda_2\mathcal{L}_{reg},
\end{equation}
where $\lambda_1$ and $\lambda_2$ are hyper-parameters balancing $\mathcal{L}_{cls}$ and $\mathcal{L}_{reg}$.

\section{Experiments}

\subsection{Experiments Settings}
\subsubsection{Datasets.}
\begin{table}[t]
    \centering
    \caption{Statistical information of experimental datasets.}
    %\vspace{-6pt}
    \label{tab1}
    %\renewcommand{\arraystretch}{1.15}
    %\resizebox{\linewidth}{!}{
    \begin{tabular}{c|ccccc}

    \toprule[1.2pt]
    Dataset                       & Domain & \#Users & \#Items & \#Interaction & Density \\ \hline
    \multirow{2}{*}{Elec\&Phone}  & Elec   & 3,325   & 17,709  & 52,966        & 0.089\% \\
                                  & Phone  & 3,325   & 38,706  & 118,114       & 0.091\% \\ \hline
    \multirow{2}{*}{Sport\&Phone} & Sport  & 4,998   & 20,845  & 54,256        & 0.052\% \\
                                  & Phone  & 4,998   & 13,655  & 46,445        & 0.068\% \\ \hline
    \multirow{2}{*}{Elec\&Cloth}  & Elec   & 15,761  & 51,447  & 224,689       & 0.027\% \\
                                  & Cloth  & 15,761  & 48,781  & 133,609       & 0.017\% \\ \toprule[1.2pt]
    \end{tabular}
    %}
\end{table}

\begin{table}[t]
\caption{The experimental results (\%) for all models, including the Recall@20 and NDCG@20 metrics. The best results are highlighted in bold, while sub-optimal results are underlined. The last row indicates the percentage improvement in performance of our method compared to the best baseline. (p-value \(<\) 0.05)}\label{tab2}
\resizebox{\linewidth}{!}{
\begin{tabular}{c|cc|cc|cc|cc|cc|cc}
\hline
Domains  & \multicolumn{2}{c|}{Elec}     & \multicolumn{2}{c|}{Phone}    & \multicolumn{2}{c|}{Sport}    & \multicolumn{2}{c|}{Phone}    & \multicolumn{2}{c|}{Elec}     & \multicolumn{2}{c}{Cloth}     \\ \hline
Metrics@20  & Recall        & NDCG          & Recall        & NDCG          & Recall        & NDCG          & Recall        & NDCG          & Recall        & NDCG          & Recall        & NDCG          \\ \hline
BPR      & 5.75          & 2.86          & 3.44          & 1.83          & 4.05          & 1.87          & 5.52          & 2.72          & 3.48          & 1.58          & 0.94          & 0.39          \\
NGCF     & 7.27          & 3.40          & 3.89          & 2.26          & 4.54          & 2.28          & 7.04          & 3.29          & 3.80          & 1.77          & 1.52          & 0.66          \\
LightGCN & {\ul 7.98}    & {\ul 3.69}    & 4.09          & 2.34          & 5.15          & {\ul 2.54}    & 7.45          & 3.48          & 3.84          & {\ul 1.82}    & {\ul 1.85}    & {\ul 0.84}    \\
DGCF     & 7.03          & 3.48          & 3.84          & 2.22          & 4.84          & 2.34          & 6.75          & 3.28          & 3.74          & 1.69          & 1.61          & 0.72          \\
MultVAE  & 6.69          & 3.24          & 3.86          & 2.28          & 4.30          & 2.02          & 6.80          & 3.10          & {\ul 3.97}    & 1.79          & 1.48          & 0.61          \\ \hline
CoNet    & 3.76          & 1.45          & 2.96          & 1.62          & 2.37          & 1.11          & 5.66          & 2.27          & 3.26          & 1.38          & 1.34          & 0.54          \\
BiTGCF   & 7.32          & 3.46          & {\ul 4.86}    & {\ul 2.79}    & {\ul 5.38}    & 2.39          & {\ul 7.83}    & {\ul 3.55}    & 3.82          & 1.71          & 1.69          & 0.76          \\
DRMTCDR  & 5.90          & 2.51          & 4.08          & 2.38          & 3.67          & 1.71          & 5.37          & 2.39          & 3.73          & 1.60          & 1.27          & 0.53          \\
DisenCDR & 5.27          & 2.12          & 3.85          & 2.18          & 3.60          & 1.59          & 6.41          & 2.67          & 3.35          & 1.42          & 1.39          & 0.55          \\ \hline
AREIL & \textbf{8.29} & \textbf{3.92} & \textbf{5.12} & \textbf{3.06} & \textbf{5.72} & \textbf{2.73} & \textbf{8.07} & \textbf{3.69} & \textbf{4.30} & \textbf{1.97} & \textbf{1.94} & \textbf{0.88} \\ \hline
Improv.  & \textbf{3.88} & \textbf{6.23} & \textbf{5.35} & \textbf{9.68} & \textbf{6.32} & \textbf{7.48} & \textbf{3.07} & \textbf{3.94} & \textbf{8.31} & \textbf{8.24} & \textbf{4.86} & \textbf{4.76} \\ \hline
\end{tabular}
}
\end{table}

\begin{table}[t]
    \centering
    \caption{Ablation study (\%) with key modules in DAFE-CDR.}\label{tab3}
    %\renewcommand{\arraystretch}{1.5}
    %\resizebox{\linewidth}{!}{
    \begin{tabular}{c|cccc|cccc}
    %\begin{tabular}{p{2.2cm}|p{1.2cm}p{1.2cm}p{1.2cm}p{1.2cm}|p{1.2cm}p{1.2cm}p{1.2cm}p{1.2cm}}
    %\begin{tabular}{p{2cm}|p{1.2cm}p{1.2cm}p{1.2cm}p{2cm}p{1.2cm}}
    \toprule[1.2pt]
    Datasets          & \multicolumn{4}{c|}{Sport\&Phone}                                                          & \multicolumn{4}{c}{Elec\&Cloth}                                                            \\ \hline
    Domain            & \multicolumn{2}{c|}{Sport}                             & \multicolumn{2}{c|}{Phone}        & \multicolumn{2}{c|}{Elec}                              & \multicolumn{2}{c}{Cloth}         \\ \hline
    Metric            & Recall          & \multicolumn{1}{c|}{NDCG}            & Recall          & NDCG            & Recall          & \multicolumn{1}{c|}{NDCG}            & Recall          & NDCG            \\ \hline
    w/o graph         & 3.91          & \multicolumn{1}{c|}{1.87}          & 5.42          & 2.70          & 3.52          & \multicolumn{1}{c|}{1.51}          & 1.34          & 0.60          \\
    w/o AREM           & 5.17          & \multicolumn{1}{c|}{2.27}          & 7.95          & 3.63          & 4.11          & \multicolumn{1}{c|}{1.92}          & 1.91          & 0.86          \\
    w/o IRLM           & 5.27          & \multicolumn{1}{c|}{2.31}          & 7.87          & 3.61          & 4.01          & \multicolumn{1}{c|}{1.88}          & 1.76          & 0.80          \\ \hline
    \textbf{AREIL} & \textbf{5.72} & \multicolumn{1}{c|}{\textbf{2.73}} & \textbf{8.07} & \textbf{3.69} & \textbf{4.30} & \multicolumn{1}{c|}{\textbf{1.97}} & \textbf{1.94} & \textbf{0.88} \\ \toprule[1.2pt]
    \end{tabular}
    %}
\end{table}

To evaluate the performance of AREIL, we conduct experiments on three real-world recommendation datasets from Amazon Datasets~\cite{Datasets}, which are widely used in cross-domain recommendation research~\cite{DisenCDR,BiTGCF,DIDA-CDR} and are considered standard benchmarks. To ensure consistency in line with previous research studies, we apply a filtering process to retain only those users who exist in both domains simultaneously. This forms three distinct cross-domain scenarios: Elec\&Phone, Sport\&Phone, and Elec\&Cloth. To convert user-item interactions into implicit data, we binarize the ratings as 0 or 1 to indicate the absence or presence of interactions. Detailed descriptions of these three cross-domain recommendation scenarios are provided in Table~\ref{tab1}.

\subsubsection{Compared Methods.}
%To evaluate the performance of AREIL, we collect a wide range of advanced methods with different views. For single-domain methods, we select traditional models such as classical matrix factorization model BPR, graph-based methods including NGCF, LightGCN and DGCF, and autoencoder-based methods such as MultVAE. Besides, we also include several widely used conventional cross-domain recommendation methods, including MLP based methods CoNet and graph-based method BiTGCF. To demonstrate the comprehensiveness of our disentanglement module, we also introduce classic disentangled cross-domain recommendation methods DRMTCDR and DisenCDR for comparison. The baselines are detailed as follows:
We compare AREIL with several classical state-of-the-art single-domain and cross-domain recommendation methods to demonstrate its effectiveness. The evaluated methods include:

\begin{itemize}
    \item[$\bullet$] \textbf{Single-domain approaches.} (1) \textbf{BPR}~\cite{BPR} utilizes collaborative filtering techniques based on matrix factorization for personalized recommendation. (2) \textbf{NGCF}~\cite{NGCF} utilizes graph convolutional networks to capture advanced collaborative signals, considering complex user-item interactions.  (3) \textbf{LightGCN}~\cite{LightGCN} improves recommendation performance by simplifying the hierarchical structure of graph convolutional networks. (4) \textbf{DGCF}~\cite{DGCF} employs dynamic graph convolutional networks and incorporates disentanglement via covariance regularization. (5) \textbf{MultVAE}~\cite{MultVAE} introduces hierarchical variational self-encoders to capture user preferences from historical interaction.
    \item[$\bullet$] \textbf{Conventional Cross-domain methods.} (1) \textbf{CoNet}~\cite{CoNet} achieves deep bidirectional knowledge migration by adding bidirectional connections to a multilayer feed-forward network. (2) \textbf{BiTGCF}~\cite{BiTGCF} realizes bidirectional knowledge transfer by exploiting the higher-order connectivity through a feature propagation layer, employing overlapped users as a bridge.
    \item[$\bullet$] \textbf{Disentanglement-based Cross-domain methods.} (1) \textbf{DRMTCDR}~\cite{DRMTCDR} disentangles user preferences into domain-shared and domain-specific components through graph contrastive learning. (2) \textbf{DisenCDR}~\cite{DisenCDR} introduces two disentanglement regularizers based on mutual information to accomplish user representation disentanglement.
\end{itemize}

%We conduct a comprehensive comparison of the proposed AREIL approach with classical and state-of-the-art single-domain and cross-domain recommendation methods. The evaluated methods include:

%These representative methods have achieved state-of-the-art results in both single-domain and cross-domain recommendation tasks, providing valuable benchmarks for comparison with the proposed AREIL method.

\subsubsection{Evaluation Protocols.}
To ensure fairness and efficiency, we employ a random partitioning strategy for each dataset, allocating 80$\%$ to the training set, 10$\%$ to the validation set, and 10$\%$ to the test set. To avoid the sampling bias of the candidate selection, we adopt the whole item set as the candidate item set during evaluation~\cite{sampled}. In this investigation, we utilize Recall (Recall@20) and Normalized Discounted Cumulative Gain (NDCG@20) as metrics for evaluating the performance of top-K recommendations~\cite{LightGCN,BiTGCF}. %Recall@K gauges the average proportion of relevant items within the top-K recommended lists, while NDCG@K assesses the positional ranking quality of the top-K lists.

\subsubsection{Parameter Settings.}
We implement the proposed AREIL method within the Recbole-CDR~\cite{Recbole} framework. During training, we set the maximum number of training rounds to 1000 and employ an early-stopping strategy based on the value of NDCG@20. For a fair comparison, we adopt the same parameter settings. The dimensions of user and item embeddings are both set to 64, while the regularization weight $\beta$ is tuned within the range of $\{1e^{-3}, 1e^{-2}, 1e^{-1}\}$. The learning rate is tuned within the range of $\{1e^{-4}, 1e^{-3}, 1e^{-2}\}$, and the Adam optimizer is employed for parameter updates. For the GNN-based model, we explore the grid search range for the number of GNN layers, considering $\{2, 3, 4\}$. 
Regarding loss weights $\lambda_1$ and $\lambda_2$, we search the values from $\{1e^{-4}, 1e^{-3}, 1e^{-2}, 1e^{-1}, 1.0\}$. Furthermore, for hyper-parameters $\gamma_s$ and $\gamma_t$ during the enhancement, we search within the range of $\{0.8, 0.85, 0.9, 0.95\}$.

% ...we employ the leave-one-out method. This approach involves selecting one sample from the dataset as the test sample and utilizing the remaining samples as training data for model training. 
% Subsequently, the trained model predicts test samples and generates a list of the top-20 recommended items. 
% Evaluation metrics such as Recall and Normalized Discounted Cumulative Gain (NDCG) are then computed. These metrics are widely used to evaluate the ranking quality of recommender systems.

\subsection{Performance Comparison}

Table~\ref{tab2} presents the experimental results of all models across the three datasets. Through the performance comparison, we can draw the following conclusions:
\begin{itemize}
    \item[$\bullet$] Cross-domain recommendation methods generally outperform single-domain recommendation methods, especially in domains with sparse data. This suggests that CDR methods effectively transfer information across domains, mitigating the issue of data sparsity.
    
    \item[$\bullet$] GNN-based recommendation methods typically yield superior performance, affirming the capability to capture higher-order collaborative information and emphasizing the necessity of intra-domain representation enhancement.
    %For example, even though LightGCN is a single-domain recommendation method, it performs well on most datasets. This validates the ability of graph-based methods to capture higher-order collaborative information and confirms the idea of employing GNNs for representation aggregation.
    %\item[$\bullet$] Decoupling-based methods (e.g., DRMTCDR, DisenCDR) do not achieve satisfactory recommendation performance, indicating that existing decoupling-based methods lack sufficient ability to achieve adaptive feature fusion and comprehensive disentanglement constraints. Our proposed method AREIL outperforms all baseline methods and achieves significant improvements across all datasets, confirming the effectiveness of our method.
    \item[$\bullet$] Disentanglement-based methods, such as DRMTCDR and DisenCDR, do not always achieve satisfactory recommendation performance on all datasets, which implies the need to develop rigorous disentanglement constraints.
    %Disentanglement-based methods, such as DRMTCDR and DisenCDR, achieve satisfactory recommendation performance in some datasets, highlighting the importance of disentangling user preferences. 
    %existing decoupling-based methods lack sufficient ability to achieve Inversed Representation Learning.
    %Our proposed AREIL method outperforms all baseline methods and achieves optimal results. It consistently shows significant improvements across all datasets.
    \item[$\bullet$] AREIL outperforms all baselines, demonstrating its effectiveness. Superiority over the GNN-based baseline confirms the value of adaptive inter-domain enhancement, while outperforming the disentanglement-based baseline validates the effectiveness of inversed learning.
     %It consistently shows significant improvements across all datasets. For instance, the NDCG@20 metric achieves a 9.68\% improvement in the Phone domain of the Elec\&Phone dataset, and a 6.23\% enhancement compared to the best baseline method in the Elec domain. These results confirm the effectiveness of our proposed method.
\end{itemize}
\begin{figure}[t]
    \centering
    \begin{subfigure}{0.43\textwidth}
        \includegraphics[width=\linewidth]{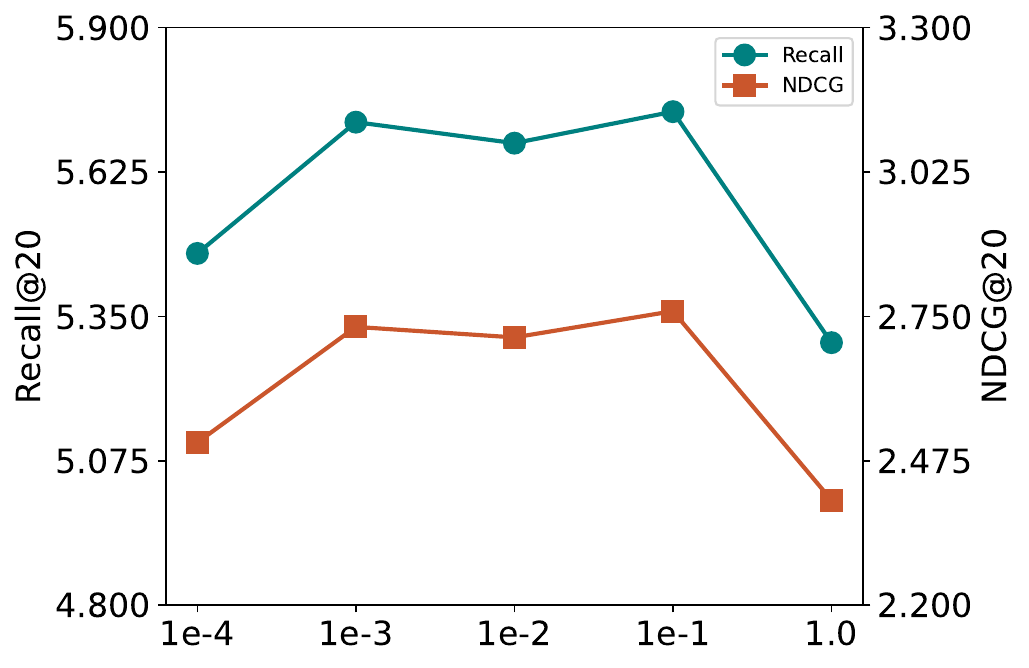}
        %\caption{Sport}
        \label{subfig:sport}
    \end{subfigure}
    \begin{subfigure}{0.42\textwidth}
        \includegraphics[width=\linewidth]{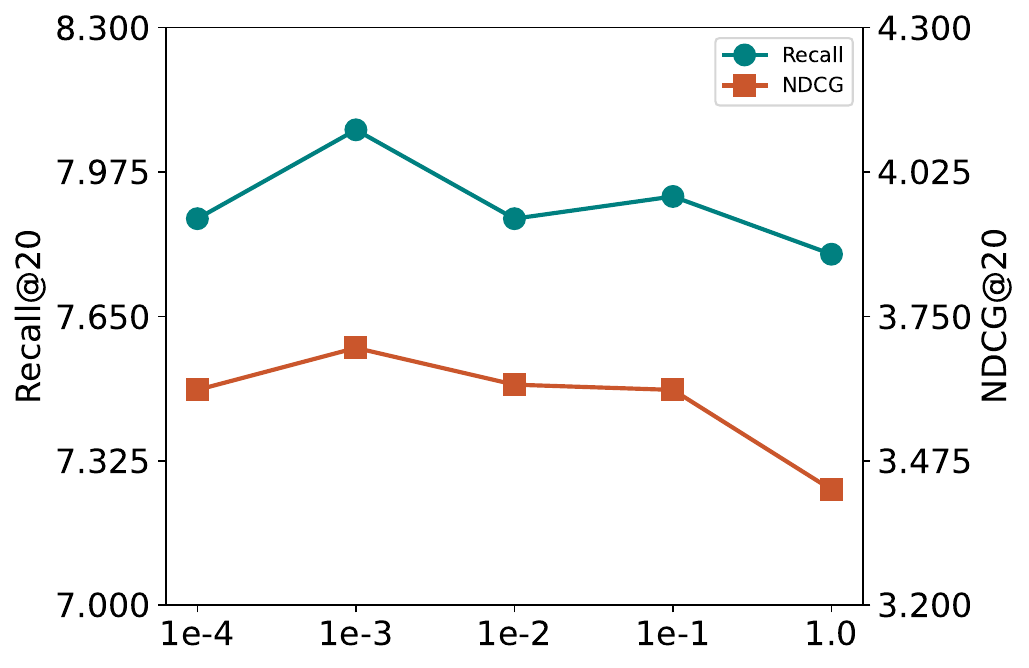}
        %\caption{Phone}
        \label{subfig:phone}
    \end{subfigure}
    \caption{Impact of the classification loss weight $\lambda_1$ in Sport(left) \& Phone(right)}
    \label{lam}
\end{figure}

\begin{figure}[t]
    \centering
    \begin{subfigure}{0.43\textwidth}
        \includegraphics[width=\linewidth]{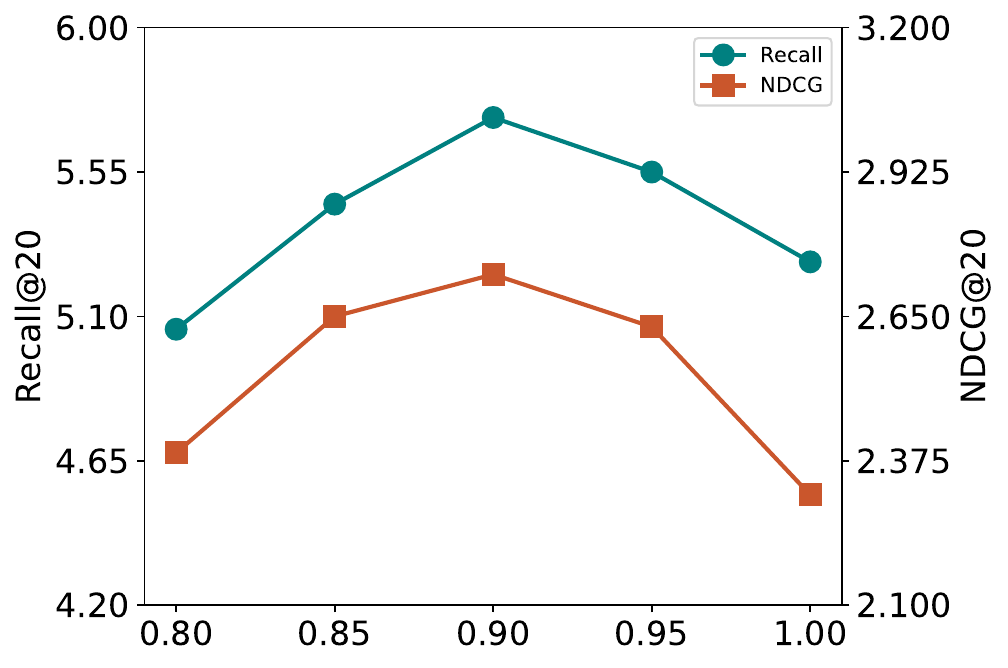}
        %\caption{Sport}
        \label{subfig:sport}
    \end{subfigure}
    \begin{subfigure}{0.43\textwidth}
        \includegraphics[width=\linewidth]{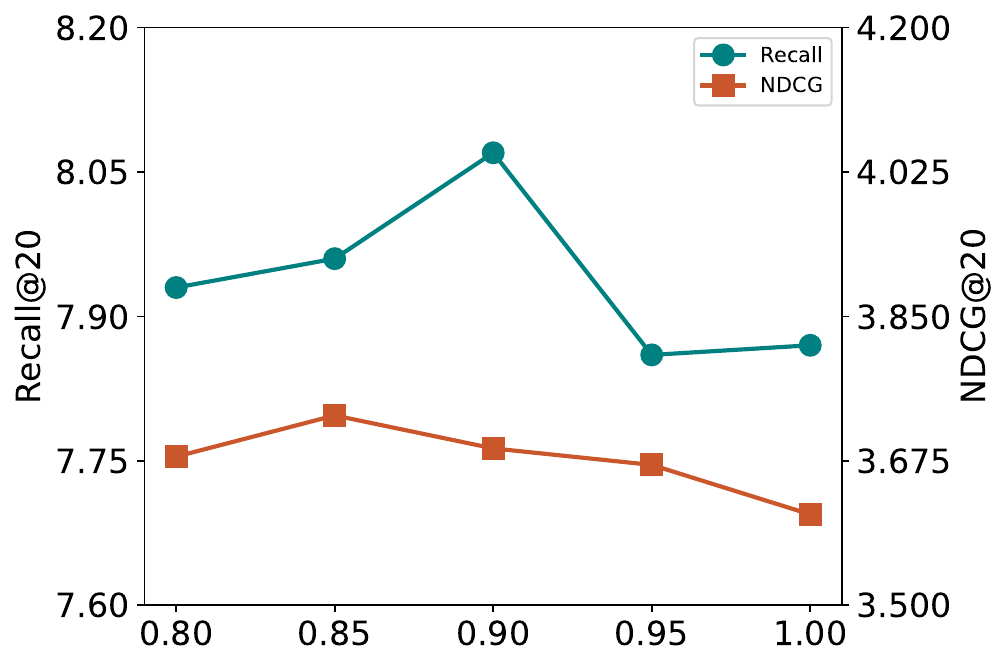}
        %\caption{Phone}
        \label{subfig:phone}
    \end{subfigure}
    \caption{Impact of the fusion controlling weight $\gamma_s$ in Sport(left) \& Phone(right)}
    \label{gams}
\end{figure}
\subsection{Ablation Study}
To validate the effectiveness of our proposed model, we conducted ablation experiments comparing AREIL with three variants: (1) \textbf{w/o graph}: replacing LightGCN in AREIL with matrix decomposition; (2) \textbf{w/o AREM}: removing the Inter-domain Adaptive Representation Enhancement Module by setting both $\gamma_s$ and $\gamma_t$ to 1; and (3) \textbf{w/o IRLM}: removing the Inversed Representation Learning Module by setting $\lambda_1$ to 0. The performance comparison between AREIL and the three variants is presented in Table~\ref{tab3}.

The results indicate that all the variants demonstrate a significant decrease in performance compared to AREIL. (1) Among them, the variant \textbf{w/o graph} exhibits the most significant performance decline. This observation implies that enhancing embeddings with intra-domain higher-order collaborative signals is not only effective but also necessary. (2) The variant \textbf{w/o AREM} demonstrates a reduction, specifically a 16.85\% decline in NDCG@20 within the Sport domain of the Sport\&Phone dataset. This underscores the inadequacy of relying solely on domain classifiers to bridge dataset disparities, and the necessity for adaptive inter-domain enhancement. (3) The variant \textbf{w/o IRLM} consistently lags behind AREIL, exhibiting a decrease of 15.38\% in the Sport domain. This emphasizes the imperative nature of inversed representation learning, a task readily achieved through harnessing the self-supervised signals provided by IRLM.

\subsection{Impact of Hyper-parameter}
%In this section, we investigate the influence of different hyper-parameters on the model performance.

\subsubsection{Impact of the classification loss weight $\lambda_1$.} %We examine the effect of the classification loss weight $\lambda_1$. Taking the Sport\&Phone dataset as an example, we adjust the range of $\lambda_1$ to $\{1e^{-4}, 1e^{-3}, \allowbreak 1e^{-2}, \allowbreak 1e^{-1}, \allowbreak 1.0\}$, while keeping the other parameters constant. The results, illustrated in Fig.~\ref{lam}, suggest that the model exhibits a limited sensitivity to the hyper-parameter $\lambda_1$ within a specific range. This diminishes the challenges associated with model tuning. A decrease in $\lambda_1$ adversely affects the model's performance as the classifier encounters difficulties, leading to suboptimal decoupling quality and a reduction in the effectiveness of the Inversed Representation Learning Module (IRLM). Conversely, an excessive $\lambda_1$ value causes the training process to deviate from the intended recommendation task, resulting in unfavorable outcomes.
The weight $\lambda_1$ of the classification loss plays a significant role in the model as it heavily influences the efficacy of inversed representation learning. Taking the Sport\&Phone dataset as an example, we vary $\lambda_1$ within the range of $\{1e^{-4}, \allowbreak 1e^{-3}, \allowbreak 1e^{-2}, \allowbreak 1e^{-1}, \allowbreak 1.0\}$, with the remaining parameters held constant. The results, depicted in Fig.~\ref{lam}, indicate that the model demonstrates constrained sensitivity to the hyper-parameter $\lambda_1$ within a defined range. The best result is achieved with a moderate $\lambda_1$ because lower values of $\lambda_1$ lead to suboptimal disentanglement quality, while excessively large $\lambda_1$ values deviate from the intended recommendation task. It demonstrates the effectiveness of our introduced domain classifiers and gradient reverse layers in inversed representation learning.

\subsubsection{Impact of the fusion controlling weight $\gamma_s$.} 
We tune $\gamma_s$ to regulate inter-domain enhancement in the source domain, exploring values within $\{0.8, 0.85, \allowbreak 0.9, 0.95, 1.0\}$. The experimental results on the Sport\&Phone dataset are presented in Fig.~\ref{gams} and optimal performance occurs when $\gamma_s$ is intermediate. If $\gamma_s$ is excessively low, the model deviates from our assumption, overlooking that features in domain-shared embeddings may vary in importance and generality across different domains. With excessively high $\gamma_s$, the model assigns low weights to the alternate domain, impeding information transfer. This demonstrates the necessity and effectiveness of the adaptive enhancement module we introduced. The trend of $\gamma_t$\ exhibits a comparable pattern. %, and we refrain from presenting intricate details.

\subsection{Representation Visualization}

\begin{figure}[t]
    \centering
    %\begin{subfigure}{0.23\textwidth}
    \begin{subfigure}{0.43\textwidth}
        \includegraphics[width=\linewidth]{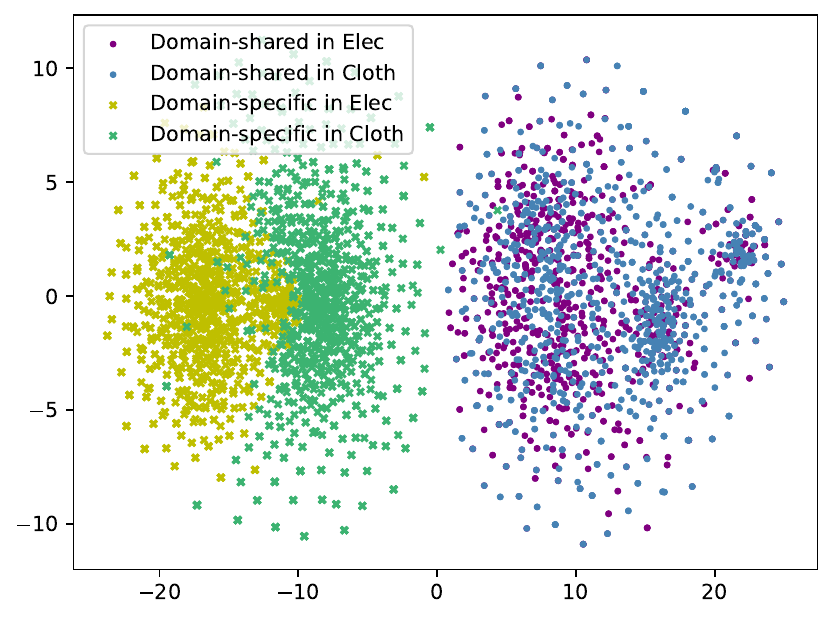}
        \caption{}
        \label{visua}
    \end{subfigure}%
    %\begin{subfigure}{0.23\textwidth}
    \begin{subfigure}{0.43\textwidth}
        \includegraphics[width=\linewidth]{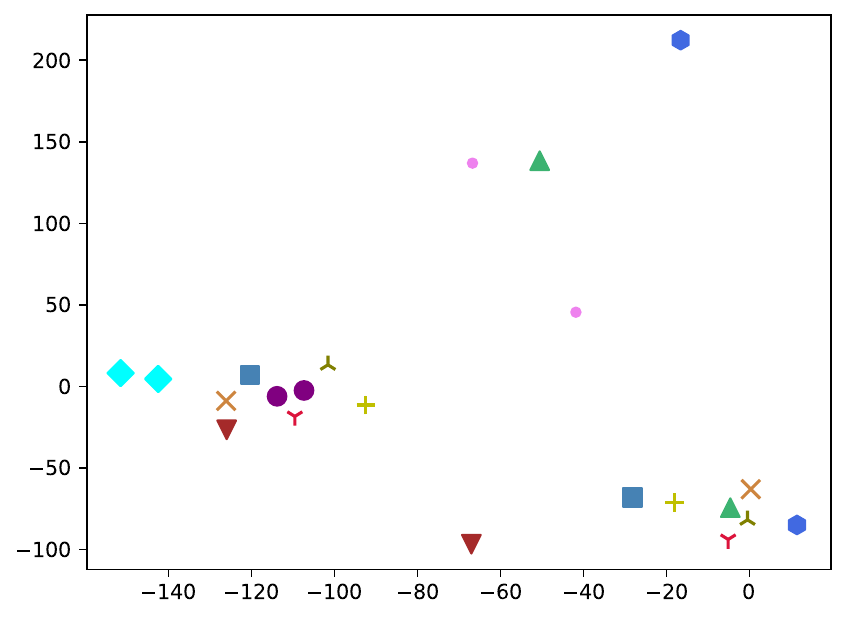}
        \caption{}
        \label{visub}
    \end{subfigure}
    \caption{Visualization of Disentangled Embeddings: (a) illustrates disentanglement in Elec\&Cloth; (b) displays the distribution of domain-shared user embeddings.}
    \label{visu}
\end{figure}

\subsubsection{Analysis on the inversed representation learning.} 
%We employ a visualization to assess the comprehensiveness of disentangled user embeddings and the necessity of achieving Adaptive Representation Enhancement across domains. Specifically, 
To provide a more intuitive display, we randomly sample 1000 pairs of disentangled user representations within the Elec\&Cloth domain, projecting them into a 2D space using t-SNE~\cite{t-SNE}, as depicted in Fig.~\ref{visua}. We can observe a distinct separation between domain-shared and domain-specific embeddings, indicating successful disentanglement of user preferences. Meanwhile, domain-shared embeddings from diverse domains overlap, while domain-specific ones maintain separation, achieving corresponding information encoding. It demonstrates that our methods provide rigorous decoupling constraints and accomplish inversed representation learning.

%We can observe that the clusters of domain-shared embeddings and domain-specific embeddings within the same domain exhibit a clear separation, indicating the disentanglement of user interests. Furthermore, domain-shared embeddings from different domains overlap, while domain-specific embeddings from different domains remain distinct, highlighting that these components encode corresponding information. We can further demonstrate that our methods achieve comprehensive disentanglement.

\subsubsection{Analysis on the necessity of adaptive enhancement.} 
In order to analyze the necessity of adaptive inter-domain enhancement, we randomly sample 10 pairs of domain-shared user embeddings and illustrate them in Fig.~\ref{visub}. Embeddings for the same user are represented by graphics with matching shapes, while distinct shapes denote different users. Notably, some embeddings demonstrate close proximity, suggesting a similarity in user preferences across domains. Meanwhile, certain pairs appear significant distance, implying disparate importance and generality in features present in both domains. Fig.~\ref{visub} underscores the imperative need for achieving Adaptive Representation Enhancement.
%We then investigate the distributions of domain-shared embeddings for the same user across different domains. To achieve this, we randomly select 10 pairs of domain-shared user embeddings. In Fig.~\ref{visub}, domain-shared embeddings for the same user in two domains are denoted by graphics of matching shapes, with distinct colors signifying different users. Notably, some embeddings exhibit close proximity, indicating a similarity in the importance and generality of these users' preferences across domains. In contrast, certain embeddings appear significantly distant from each other, suggesting that features existing in both domains exhibit disparate behavior. This observation expresses the necessity for designing Adaptive Representation Enhancement.%This observation confirms our motivation for feature fusion and expresses the necessity for designing Adaptive Representation Enhancement.

\section{Conclusion}

This paper introduced a novel approach to dual-target cross-domain recommendation by focusing on adaptive representation enhancement and inversed representation learning. Specifically, we first disentangled mixed user preferences by dividing user representations into domain-shared and domain-specific components. To further improve the ability of user representation, we proposed an adaptive enhancement module that captured high-order information and revealed inter-domain correlations. Next, within a unified framework, we leveraged inversed constraints to learn truly disentangled user preferences. At last, we optimized the entire framework via multi-task learning. Extensive experiments demonstrate that AREIL significantly outperforms state-of-the-art baselines. In the future, we will explore incorporating additional attribute information for even more efficient inter-domain enhancement.

\bibliographystyle{splncs04}
\bibliography{Sections/References}
\end{document}